\documentclass[11pt,twoside]{article}

\usepackage{asp2004}
\usepackage{epsf}
\usepackage{psfig}
\usepackage{lscape}

\begin{document}

\vspace*{2em}

\title{Steps Toward a Common Near-Infrared Photometric System}  

\author{A. T. Tokunaga}   
\affil{Institute for Astronomy, University of Hawaii, 2680 Woodlawn Dr., 
Honolulu, HI  96822}    
\author{W. D. Vacca}   
\affil{Stratospheric Observatory for Infrared Astronomy/Universities Space 
Research Association, NASA Ames Research Center, MS N211-3, 
Moffett Field, CA 94035-1000}

\begin{abstract} 
The proliferation of near-infrared (1--5 $\mu$m) photometric systems
over the last 30 years has made the comparison of photometric results
difficult.  In an effort to standardize infrared filters in use, the
Mauna Kea Observatories near-infrared filter set has been promoted among 
instrument groups  through combined filter production runs.  The characteristics 
of this filter set are summarized, and some aspects of the filter
wavelength definitions, the flux density for zero magnitude, atmospheric extinction
coefficients, and color correction to above the atmosphere are discussed.

\end{abstract}

\section{Introduction}   

The foundations of modern infrared photometry at near-infrared
wavelengths (1-5 $\mu$m) were built upon the pioneering work of
Johnson and his collaborators  \citep{joh66a,joh66b,joh68}.  
The near-infrared filters used in these
early works were very broad, and the atmospheric absorption bands
defined the effective widths of the filters. The filter profiles used by
\citet{joh65} are shown in Figure 1, which demonstrates that
the deep atmospheric absorption bands (mostly due to water vapor) play
a significant role in determining the effective transmission of the
atmosphere + telescope + filter observational system. 

The problem with
allowing the atmosphere to define the passband is that the atmosphere
varies substantially from place to place and, at any given location,
from night to night (even from hour to hour). Hence, the reproducibility of
photometric observations  and conversions to other photometric systems
becomes difficult or impossible to achieve. To prevent these problems,
the photometric bands should ideally be defined by only the filter
transmission profiles (if possible).  Ideally, the filter bandpasses should avoid 
deep telluric absorption features and any part of the telluric spectrum
that has a large first derivative and/or is highly variable. In
principle, filter properties can be controlled by
the manufacturer, and so the specifications and characteristics of the 
filters are usually the only aspect of the observational system that
can remain constant from instrument to instrument and site to site. If
such a set of filters were easily replicable and distributed to a
large number of observatories, the comparison of photometric results
from various instruments, telescopes, and sites would be much more
straightforward.

There has never been an agreed upon filter set for the ground-based
near-infrared filters.  As a result, many types of filters are
currently used, as illustrated in Figure 2, and summaries of infrared
photometric systems in use are given by  
\citet{bes05}, \citet{gla99}, and \citet{tok00}.  (The former two references
also provide excellent introductions to infrared photometry.) As a
result of this proliferation of filter sets, the comparison of results
in different systems can be quite complicated. Careful attention to
the details of the photometric system employed  such as that
provided by \citet{mor00} and \citet{fio03} is needed. 
Furthermore, the labels for the filters in different sets are often
the same (e.g., $J$, $H$, and $K$), and this can lead the
unwary to mistakenly combine or compare results inappropriately. Very
different filters may have been used for observations despite the
common label, and corrections for the differing filter (isophotal)
wavelengths must be incorporated first before comparisons can be
done.  For objects with very deep absorption bands or strong emission
lines, direct comparison may not be possible. 

The basic difficulty of having different filters lies in the intercomparison of
results from the various photometric systems. In
order to make accurate comparisons, all of the photometric points must
be either (1) plotted at the filter wavelengths
of the various systems, or (2) corrected to the same wavelength. The
latter is the purpose behind color transformations.
Intercomparison of results among the different systems requires color
transformations such as those provided by 
\citet{car01}, \citet{haw01}, \citet{ste04}, and \citet{leg06}.

The large number of color transformations highlights the need for
standardization of infrared filters. As stated in the goals for this workshop:
``The astronomical community witnesses an explosive non-linear growth
of observational capability \dots This quantum leap and its
associated uncontrolled production of diverse systems of measurement
disrupts the consistency of calibrated measurements \dots ''.  Figure 2
serves as a fitting testimony to the relevance of these statements.  

Although color transformations are very helpful in combining data from
different systems, the systematic effects and limited color range of
the transformations make it much less desirable than having
observations with a single filter set. The adoption of a standard
filter set would minimize confusion about which filter set is being
used and the magnitude of the color transformations. However, use of a
single filter set will not eliminate the need for color
transformations due to optical efficiency differences among
instruments and because completely identical filters cannot be
fabricated for all photometric systems.

A relatively new set of filters, the Mauna Kea Observatories
near-infrared (MKO-NIR) filter set, was designed for minimizing the
atmospheric absorption effects.   The filter profiles are shown in
Figure 1. These filters are  located near the center of
the high atmospheric transmission regions, and therefore the filter central
wavelengths are defined by these regions.

We present here a brief summary of the properties of the MKO-NIR filter set.  
This filter set is gaining acceptance in the community and thus could serve
as a unified near-infrared filter set in the future.  We also discuss
aspects of infrared photometry that may be helpful to newcomers to 
infrared photometry, and we discuss a method to establish an improved 
photometric system for the MKO-NIR filter set.

\section{The MKO-NIR Filter Set} 

A filter set for 1--5 $\mu$m was developed initially for the Subaru
and Gemini telescopes.  The central wavelength and bandpass were
optimized by maximizing throughput while minimizing the effects of the
atmospheric transmission and emission.  The method to determine this
was presented by \citet{sim02}.  The 1996--2002 period was
an opportune time to convince many other groups to use these filters, since
there were a large number of new infrared instruments under
development at the time. There was a strong motivation to combine
orders for production, thus sharing the engineering and fabrication costs
and thus greatly reducing the cost per filter. In
addition, the filters were optimized at both 2 and 4 km altitude, and
thus many observatories could use these filters. As a
result these filters are now widely used at a large number of major 
observatories and are being employed in near-infrared sky surveys 
such as the UKIRT Infrared Deep Sky Survey (UKIDSS).

The detailed specifications for the fabrication of the filters and 
atmospheric extinction coefficients are described by \citet{tok02}, 
while the isophotol wavelengths (for a Vega source spectrum) and absolute 
calibration are described by \citet{tok05}.  

\subsection{Definition of the Filter Wavelengths}

There is no unique definition of the ``effective wavelength'' of a broad-band
filter.  See \citet{gol74} and \citet{tok05} for detailed discussions of the various
filter definitions. The fundamental problem is that the filters are broad
relative to the wavelength scale of the variations in the source spectra. 
In contrast a single pixel in a spectrograph 
is like a filter with a very well defined wavelength simply because
its wavelength span is small compared with the variations in the spectra
in most sources.

As we show below (see also \citealt{gol74}), the isophotal wavelength is the
most appropriate filter wavelength definition to use. This definition is {\em not} 
independent of the source spectrum, which presents problems, and is
the reason others have used different filter wavelength definitions
in an attempt to approximate the isophotal wavelength without having
to incorporate the source spectrum.

To derive an expression for the isophotal wavelength we first consider
the number of  photo-electrons detected per second from a
source with an intrinsic spectral energy distribution
$F_\lambda(\lambda)$.  This is given by
\begin{eqnarray}
\label{eq:pdef}
N_p & = & \int F_\lambda(\lambda) S(\lambda)/ h\nu \ d\lambda \\
\label{eq:fdef}
         & = & \frac{1}{hc}\int \lambda F_\lambda(\lambda) S(\lambda) d\lambda ~~,
\end{eqnarray}
where  $S(\lambda)$ is  the total system response given by 
\begin{equation}
S(\lambda) = T(\lambda) Q(\lambda) R(\lambda)  A_{\rm{tel}}  ~~.
\end{equation}  
Here $T(\lambda)$ is the atmospheric transmission, $Q(\lambda)$ is the
product of the throughput of the telescope, instrument, and quantum
efficiency of the detector, $R(\lambda)$ is the filter response
function, and $A_{\rm{tel}}$ is the telescope collecting area.  The
system response $S(\lambda)$ is equal to the relative spectral
response (RSR) defined by \citet{coh03a}. Care is needed when
taking filter response functions from various sources. For example,
the 2MASS filter response functions have already been corrected for
the extra factor of $\lambda$ in the integral \citep{coh03b}.

If $F_\lambda(\lambda)$ and $S(\lambda)$ are both continuous, and
$S(\lambda)$ is not negative over the wavelength interval, then from
equation (2) and the mean value theorem for integration there exists
a wavelength, $\lambda_{\rm{iso}}$,  such that 
\begin{equation}
F_\lambda(\lambda_{\rm{iso}})   \int \lambda S(\lambda) d\lambda  = 
\int \lambda  F_\lambda(\lambda)  S(\lambda) d\lambda  ~~.
\end{equation}
Rearranging this, we obtain
\begin{equation}
F_\lambda(\lambda_{\rm{iso}})  =  \langle F_\lambda \rangle  =  
\frac{\int \lambda F_\lambda(\lambda) S(\lambda) d\lambda} {\int \lambda S(\lambda) d\lambda} ~~,
\end{equation}
where $\lambda_{\rm{iso}}$ is known as the ``isophotal wavelength" and
$\langle F_\lambda \rangle$ denotes the mean value of the intrinsic
flux above the atmosphere (in units  of W m$^{-2}$ $\mu$m$^{-1}$) over
the wavelength interval of the filter. 

In a similar fashion,
\begin{equation}
F_\nu(\nu_{\rm{iso}})  =  \langle F_\nu \rangle  =  
\frac{ \int  F_\nu(\nu) S(\nu)/\nu \ d\nu} {\int S(\nu)/\nu \ d\nu} ~~,
\end{equation}
where $\nu_{\rm{iso}}$ denotes the ``isophotal frequency'' and
$\langle F_\nu \rangle$ denotes the mean value of the intrinsic flux
above the atmosphere (in units W m$^{-2}$ Hz$^{-1}$) over the
frequency interval of the filter.  

The isophotal wavelengths and frequencies, $\lambda_{\rm{iso}}$ and
$\nu_{\rm{iso}}$, respectively, for the MKO-NIR filters are given in
Table 1 (see also \citealt{tok05}). These were computed
using the model atmosphere for Vega recommended by \citet{boh04} 
and therefore strictly apply only to A0V stars. For any other
source spectrum, the isophotal wavelengths would differ from those
shown in Table 1.

The isophotal wavelength is important and useful because it is
the wavelength at which the monochromatic flux $F_\lambda $ equals the
mean flux in the passband. As stated succinctly by \citet[][p. 41]{gol74}: 
``The wavelength of the monochromatic intensity deduced from a
heterochromatic measurement  \dots  is  \ldots   $\lambda_{iso}$.''
The use of isophotal wavelengths determined from a model of Vega
provides consistency with the extensive series of papers on infrared
calibration by Cohen and his collaborators. Other definitions for the
filter wavelength are discussed by \citet{gol74} and \citet{tok05}.

We note that the UKIDSS program defined filters for $Z$ and $Y$ with
effective wavelengths at 0.88 $\mu$m  and 1.03 $\mu$m respectively, 
\citep{hew06}.  It is typically the case that the longest
filter for CCD work is named ``$Z$''.  The ``$Y$'' filter was introduced
by \citet{hil02} to take advantage of the atmospheric
window between atmospheric absorption at 0.955 and 1.112 $\mu$m. The
definition of the effective wavelength adopted by the Sloan Digital
Sky Survey  as well as the UKIDSS \citep{hew06} is given by
 \citep[][see their equation 3]{fuk96} and is not that of the
isophotal wavelength for an A0V star. Therefore, the flux densities in
Hewett et al.'s Table 7, which correspond to the flux densities of an
object of zero magnitude of constant flux density over the passband,
are not the monochromatic flux densities at the effective wavelengths
given in their table. Thus, while the flux density zero points agree
well with those given in Table 1, direct comparison of the
wavelengths in Table 1 with that of Hewett et al.'s Table 7 is
not appropriate.

\subsection{Correction for Spectral Energy Distribution}

Photometric data points are often converted to flux densities and
interpreted as measurements of the monochromatic flux at wavelengths
given by the ``effective wavelengths'' of the filters, such as the
isophotal wavelengths.  However, what one has measured is the
``in-band'' flux passing through the filter.  Since photometry is
unavoidably a heterochromatic (multiwavelength) measurement,
observations of sources with very different spectral energy
distributions cannot be directly compared at a single wavelength
unless a correction is made for the different spectral energy
distributions.  This is clear from the definition of the isophotal
wavelength, where the spectral energy distribution must be known to
compute it.

The dependence of the isophotal wavelength on the source spectrum
is the fundamental reason for the difficulties encountered in the
interpretation and intercomparison of photometric results. To accurately
know the wavelength at which a photometric data point
applies, one needs to know the source spectrum, which is usually what
one is attempting to determine in the first place. The application of
a correction factor during photometric reductions is a means of 
attempting to account for
the sensitivity of $\lambda_{\rm {iso}}$ to the slope of the source
spectrum, and thereby estimating the monochromatic source flux density at the
isophotal wavelength of an A0V star.

To visualize this, consider the observation of a very cool object, say
$T \sim 500$~K.  Suppose the standard star is much hotter, 
$T \sim 10^4$ K.  Then the isophotal wavelength of the standard is
inappropriate for the cool object since the flux density is weighted
to the longer wavelengths compared to the standard star.  If a
correction is not made for the spectral energy distribution, then one
will overestimate the flux density of the cool object at the isophotal
wavelength of the standard.  To obtain the proper flux density of the
cool object at the isophotal wavelength of the standard one must apply
a correction factor that reduces the flux density of the cool object.  These
correction factors are required for any definition of the filter wavelength,
including those that are independent of the source spectrum. 
Discussions of computing the correction factor can be found in 
\citet{low74}, \citet{han84}, and \citet{gla99}.

\subsection{Conversion of Magnitudes to Flux Density}

All photometry is relative, and so it is necessary to define the
photometric zero point by using a star or group of stars to define
zero magnitude and zero colors. In establishing the $UBV$ system, 
\citet{joh53} tied the $V$ magnitude to nine stars of the
International Photovisual system and the $U-B$ and $B-V$ colors to an
average of six A0V stars.  Most optical photometric systems assume a 
visual magnitude of $+$0.03 for Vega.  However most infrared 
photometric systems, such as
\citet{eli82} for the C.I.T. system and  \citet{coh92} for
his absolute calibration work, define Vega to have zero magnitude 
and zero color at 1--5 $\mu$m.  More recently \citet{pri04a} have
chosen to tie the MSX photometry to Sirius since Vega has dust
emission that affects its brightness at wavelengths longer than 12
$\mu$m.
Therefore, in comparing near-infrared photometric systems, care is
required to make sure that the assumptions about the visual and 
infrared magnitude of Vega are consistent. 
\citet{coh92} assumed 0.0 mag for Vega in the near-infrared, but in 
\citep{coh03a} the nonzero magnitude of Vega at $V$ was taken into
account in establishing the 0.0 mag flux densities (see their 
Table 3).

Vega is the most well-studied star for the determination of the
absolute flux density.  A summary of the direct comparison of blackbody 
furnaces to Vega is given by  \citet{meg95}, while \citet{coh92} 
used a model atmosphere to extrapolate from the absolute
calibration at $V$ to the near infrared.  \citet{tok05} show
that these two methods give the identical flux density in the
near-infrared.

The flux density for Vega shown in Table 1 is from \citet{tok05}, 
and it is normalized to be consistent with the flux
density of Vega determined by \citet{coh92}.   \citet{tok05} 
have shown that these values are consistent with the
model-independent absolute calibration measurements summarized by
\citet{meg95}.  In addition, the absolute calibration of \citet{coh92} 
was found to be consistent with the extensive absolute
calibration conducted by the {\em Midcourse Space Experiment} 
\citep{pri04a,pri04b} to within 1\%.   Therefore the values shown in 
Table 1 are consistent with the extensive absolute calibration work 
of Cohen and his collaborators, of which \citet{coh03b} is the 
latest in the series.

Nevertheless, the use of a model atmosphere for Vega is fraught with
difficulties because Vega is a rapid rotator and is
observed nearly pole-on \linebreak  \citep{boh04,pet06}.   Consequently
there is a temperature gradient across the
surface of Vega as viewed from Earth.  Therefore, a model
atmosphere with a single fixed temperature cannot provide an accurate
representation of the observed flux density of Vega from the
ultraviolet to the near-infrared.  In fact, there is no self-consistent
model of Vega that is applicable from the optical to the infrared. 
Nonetheless, the single temperature model assumed by \citet{coh92} 
provides a close approximation to the near-infrared spectral
energy distribution that we observe and has been validated by
ground-based and space-based absolute calibration experiments.  This
does not imply the model atmosphere is a valid one but rather that the
model atmosphere is empirically a good match to the near-infrared
spectral energy distribution of Vega.

\subsection{Linearity and Extinction}

In establishing the $UBV$ system \citet{joh53} employed the technique
of linearly extrapolating magnitudes to above the atmosphere.  A strictly 
linear extrapolation is not valid due to the Forbes effect; the atmospheric
extinction curve is nonlinear between zero and unity airmass.  This is
shown explicitly for the MKO-NIR filters in Figure 3.  The nonlinearity of the 
extinction in the near-infrared passbands was discussed by 
\citep[][see their Fig. 2]{man79}, \citet{you74}, and \citet{you94} 
and arises from the very strong
absorption bands of water vapor in the near-infrared.   As a result, all
commonly used near-infrared photometric systems (Glass, Elias,
Persson, UKIRT) employ the method of reducing to an airmass of 1.0. 
This is acceptable since all photometry is done differentially.

The effects of water vapor absorption are not 
completely eliminated by the design of the MKO-NIR filters.  
\citet{tok02} computed the extinction for the MKO-NIR filters as a 
function of telluric precipitable water.  The atmospheric extinction 
coefficients is shown in Table 2 for 2 mm of precipitable water. 
For $J$, $H$, and $K$ these values are smaller than for those 
given by \citet{kri87} who employed older filters with wider passbands.   
See also the UKIRT Web site,  
http://www.jach.hawaii.edu/UKIRT/astronomy/ utils/exts.html).  
For $L'$ and $M'$ the calculated extinction coefficients are identical 
to that observed within the errors.

Computations of the extrapolation to zero airmass as a function of
water vapor from 1 to 5 mm precipitable water suggests the variations
of the magnitude at zero airmass is in the range of 0.003--0.012 mag 
for $J$, $H$, $K$, $K'$, $K_{\rm{s}}$, and $L'$, and in the range of 0.096-0.123
for $M'$ for the MKO-NIR filters \citep{tok05}. For nearly
all work the uncertainties in the exo-atmospheric magnitudes for $J$, $H$,
$K$, and $L'$ are acceptable. Since the uncertainties at $M'$ are
larger, care is needed if accurate exo-atmospheric magnitudes are
needed.

A complication arises for very cool stars that have deep water and
other strong molecular absorption bands.  In this case, there will be a
color term in the extinction curve.  Good discussions on measuring the
color term are given by \citet{har62} and  \citet{hen82}, 
and the formal derivation of the color term in photometry is given by
\citet{gol74}. As this color term is rarely measured,
high-precision photometry of cool objects is subject to uncertainties
arising from neglecting it.
The importance of this color term is that when one extrapolates the
observations to zero airmass (i.e., above the atmosphere), the result
should be identical to observations done completely in the absence of
an atmosphere.  

\citet{mil05} have proposed relatively narrowband filters
that minimize color transformations between observatories and
instruments.   The extrapolation to zero airmass for these filters 
is approximately
linear.  This filter set is important for observations where one
wants to determine the absolute calibration to above the atmosphere.   
The absolute calibration work by \citet{bla83} and
\citet{sel83} used narrowband filters exactly for this reason.

\section{Setting the Photometric Zero Point for the MKO-NIR Filters}

The papers by \citet{haw01}, \citet{leg03}, and \citet{leg06} give a set of
standard stars measured in the MKO-NIR filter set.  The magnitudes of
these standard stars were largely based on the system established by
\citet{eli82}.  Although the zero point of the system was not
determined explicitly,  it should be close to that of Elias, namely that the
colors of Vega are 0.0 mag
($V-J=J-H=H-K=K-L'=L'-M'=0.0$).  We discuss in this
section an approach to better establish the photometric zero point of
the MKO-NIR standards relative to a set of A0V stars.

To set the zero point of the MKO-NIR photometric system,
one could observe a number of A0V stars and either define Sirius to
have zero colors as suggested by \citet{coh92} and \citet{pri04a}
or adopt an average of A0V stars to have zero colors after
removing the interstellar extinction.  The latter method was used by
\citet{joh53} when they defined the $UBV$ system.  Thus Vega has a
magnitude of +0.03 mag in the Johnson system, although 
\citet{boh04} recently determined $V=+0.026 \pm 0.008$ mag.  In
the former method, Sirius is proposed instead of Vega because there is
no definitive model atmosphere for Vega and because Sirius is accessible to the
Northern and Southern hemispheres.  However Sirius and Vega are too
bright to be observed with current instrumentation.

\citet{hew06} have computed the colors of stars in all spectral classes using 
synthetic photometry (see \citealt{bes05} for a description of this technique). 
To set the zero point, they used the Vega atmospheric model adopted by
\citet{boh04} and assumed that the magnitude of Vega is
zero at all wavelegths.  This approach has the advantage that
corrections can be easily made for adoption of other magnitudes for
Vega or an improved atmospheric model for Vega.

\begin{sloppypar}
\citet{hew06} used an inhomogeneous set of sources in the
literature for the spectra of the M, L, and T dwarfs. There is
currently a program underway at the IRTF to generate a library of
moderate resolution, high signal-to-noise spectra of a large number of
stars with reliable spectral types spanning the entire range of known
spectral types and luminosity classes. Spectra for M, L, and T dwarfs,
included as part of this IRTF spectral library, have already been
published \citep{cus05} and spectra for F--T stars are now
available on the IRTF Web site
http://irtfweb.ifa.hawaii.edu/$\sim$spex/spexlibrary/ IRTFlibrary.html. We
are undertaking a project to use these spectra to compute synthetic
photometry of the stars in the library  to significantly
improve on the approach taken by \citet{hew06}. The advantages
of using this spectral library are  (1) it covers the spectral range
0.8--5.0 $\mu$m, with resolutions of $R=2000$ between 0.8--2.5~$\mu$m,
and $R=2500$ between 2.5--5.0~$\mu$m, and signal-to-noise ratios
greater than 100; (2) all of the data were obtained with the same
instrument (SpeX, \citealt{ray03}) and reduced in a consistent
manner with the same reduction package (Spextool, \citealt{cus04}); 
(3) all flux calibrations were achieved using A0V stars, with $V$
mag between 5--8, as spectral standards \citep{vac03}; and (4)
most of the stars in the spectral library have 2MASS magnitudes
available.
\end{sloppypar}

Because the IRTF spectral library calibration relies on the model of
Vega adopted by  \citet{tok05} and observations of A0V
stars, scaled to their observed $V$ mag and corrected for any possible
reddening as estimated from their $B-V$ colors  \citep[see][]{vac03},
the colors of the program stars as derived from synthetic photometry, 
should be quite accurate.  
To determine the systematic uncertainties, we plan to obtain
photometry on a set of unreddened A0V stars with the MKO-NIR filters.
The average color of these stars would be defined to be zero for all
filter combinations.  These stars would be then observed in a 
fashion similar to that used for the IRTF spectral library and compared to the
photometric results.  Previous comparisons between synthetic photometry
and actual measurements indicate that the
uncertainties in the colors are at the level of a few percent \citep{cus04}.

\section{Summary}

Infrared photometry is now a mature technique with photometric
accuracy comparable to that obtained in the visible wavelength range. 
However, standardization has not yet been achieved. Combining data
from different observations is generally difficult due to the diverse filter 
sets (often with similar labels) and the numerous photometric
systems now in use.  We advocate the use of a single filter set, and
the widespread use of the MKO-NIR filter set is helping to achieve this. 
We have summarized some aspects of infrared photometry for the 
novice to infrared photometry-- filter wavelength definitions, the flux 
density for zero magnitude, atmospheric extinction coefficients, and color
corrections to above the atmosphere.

\acknowledgements 
A.\ T.\ T.\ acknowledges the support of NASA Cooperative Agreement NCC 5-538.
W.\ D.\ V.\ acknowledges partial support from Prof.\ James R.\ Graham.

\section{APPENDIX A}

Web-based photometric information and color transformations can be found at
the following locations: \\

\noindent Asiago Database on Photometric Systems:  \\
\indent  http://ulisse.pd.astro.it/Astro/ADPS/Systems/index.html  \\
Catalog of Infrared Observations: \\
\indent  http://ircatalog.gsfc.nasa.gov/ \\
Gemini Near-Infrared Photometric Standard Stars: \\
\indent  http://www.gemini.edu/sciops/instruments/niri/NIRIPhotStandards.html \\
2MASS color transformations: \\
\indent  http://www.astro.caltech.edu/$\sim$jmc/2mass/v3/transformations/ \\
UKIRT Photometric Calibration:  \\
\indent  http://www.jach.hawaii.edu/UKIRT/astronomy/calib/phot\_cal/ \\
IRTF photometric information: \\
\indent  http://irtfweb.ifa.hawaii.edu/IRrefdata/ph\_catalogs.html; \\
The MKO-NIR filter profiles can be obtained at:  \\
\indent http://irtfweb.ifa.hawaii.edu/$\sim$nsfcam/hist/filters.2006.html \\


\clearpage

\begin{table}[!ht]
\caption{Isophotal Wavelengths and Absolute Flux Densities for Vega}
\smallskip
\begin{center}
{\small 

\begin{tabular}{ccccc}
\tableline
\noalign{\smallskip}

Filter & $\lambda_{\rm{iso}}$ & F$_\lambda$   & $\nu_{\rm{iso}}$   & F$_\nu$  \\
   & ($\mu$m) & ($\times$10$^{-11} $W m$^{-2}$ $\mu$m$^{-1}$)  & ($\times$10$^{14}$ Hz) & (Jy)  \\
\tableline

$V$	      &0.5450	&3680.  &  5.490   &  3630     \\
$J$	      &1.250	&301.	  & 2.394    &  1560     \\
$H$	      &1.644	&118.	 &  1.802    &  1050     \\
$K'$		&2.121	&45.7   &  1.413    &  686      \\
$K_{\rm{s}}$	      &2.149	&43.5	 &  1.395    &  670      \\
$K$	      &2.198	&40.0	 &  1.364    &  645      \\
$L'$		&3.754	&5.31	 &  0.7982  &  249     \\
$M'$		&4.702	&2.22	 &  0.6350  &  163    \\

\noalign{\smallskip}
\tableline
\end{tabular}
}
\end{center}
\end{table}
Notes: (1) From \citet{tok05}.  The uncertainty in the absolute flux density is 1.5\%.
(2) $\lambda_{\rm{iso}}$ and $\nu_{\rm{iso}}$ were computed from using equations 
(5) and (6).  Due to the wide bandpasses of the filters, 
$\lambda_{\rm{iso}}$ $\neq$ c/$\nu_{\rm{iso}}$  and 
F$_\nu$(Jy) $\neq$ ($\lambda_{\rm{iso}}$($\mu$m))$^2$ $\times$ 
F$_\lambda$(W m$^{-2}$ $\mu$m$^{-1}$)/(3$\times$10$^{-12}$).

\clearpage

\begin{table}[!ht]
\caption{Linear Fit to Calculated Extinction Curve for 2 mm H$_{2}$O}
\smallskip
\begin{center}
{\small 

\begin{tabular}{ccc}
\tableline
\noalign{\smallskip}

        & Constant   & Slope  \\
Filter  & (mag)      &  (mag/airmass)  \\

\tableline

$J$         &  0.0085   $\pm$ 0.0005   &  0.0153   $\pm$ 0.0003  \\
$H$        &  0.0091   $\pm$ 0.0005    & 0.0149   $\pm$ 0.0003 \\
$K$        & 0.0175   $\pm$ 0.0008    & 0.0331   $\pm$ 0.0005 \\
$K'$   & 0.0731   $\pm$ 0.0027  &   0.0589   $\pm$ 0.0015  \\
$K_s$ & 0.0442   $\pm$ 0.0017   &  0.0429   $\pm$ 0.0010  \\
$L'$    & 0.0264   $\pm$ 0.0020   &  0.1039   $\pm$ 0.0012  \\
$M'$  & 0.1099   $\pm$ 0.0049  &   0.2226   $\pm$ 0.0028 \\

\noalign{\smallskip}
\tableline
\end{tabular}
}
\end{center}
\end{table}

Notes: 
(1) From \citet{tok02}.
(2) Least-squares linear fitting to the equation 
(const. + slope*X), where X is the airmass in the range 1.0--3.0.  

\clearpage

\begin{figure}
\plotone{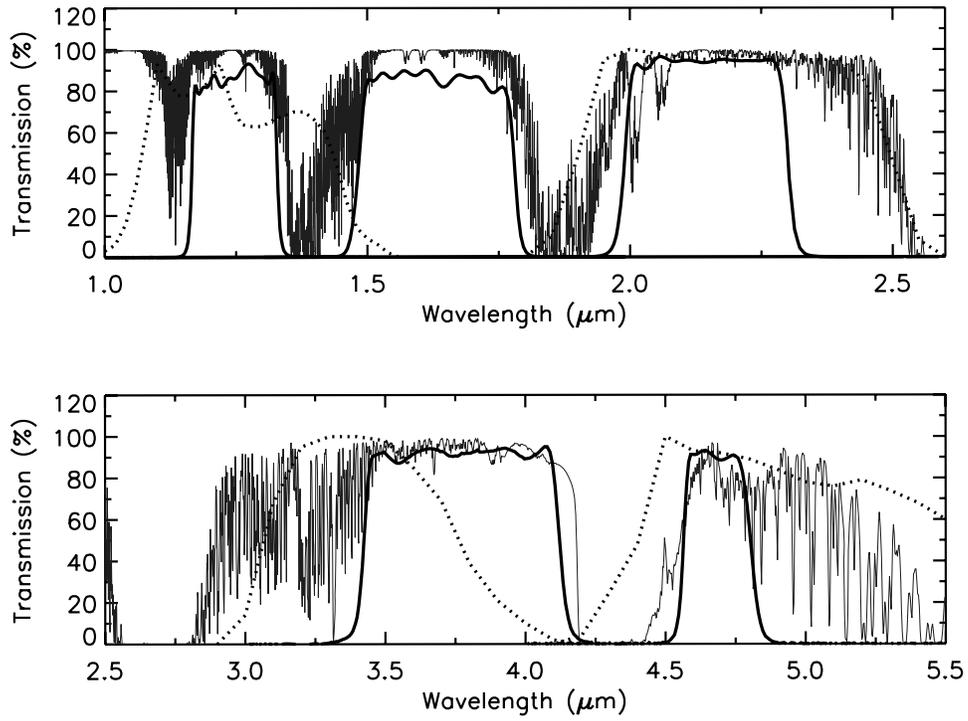}
\caption{ Comparison of the \citet{joh65} filters (dotted lines) with
the MKO-NIR filters (solid lines).  Note that the original
Johnson infrared filter set did not include the H band.  The
atmospheric transmission for Mauna Kea with 1 mm precipitable water
and an airmass of 1.0  is shown for comparison.
}
\end{figure}

\clearpage
\begin{figure}
\plotone{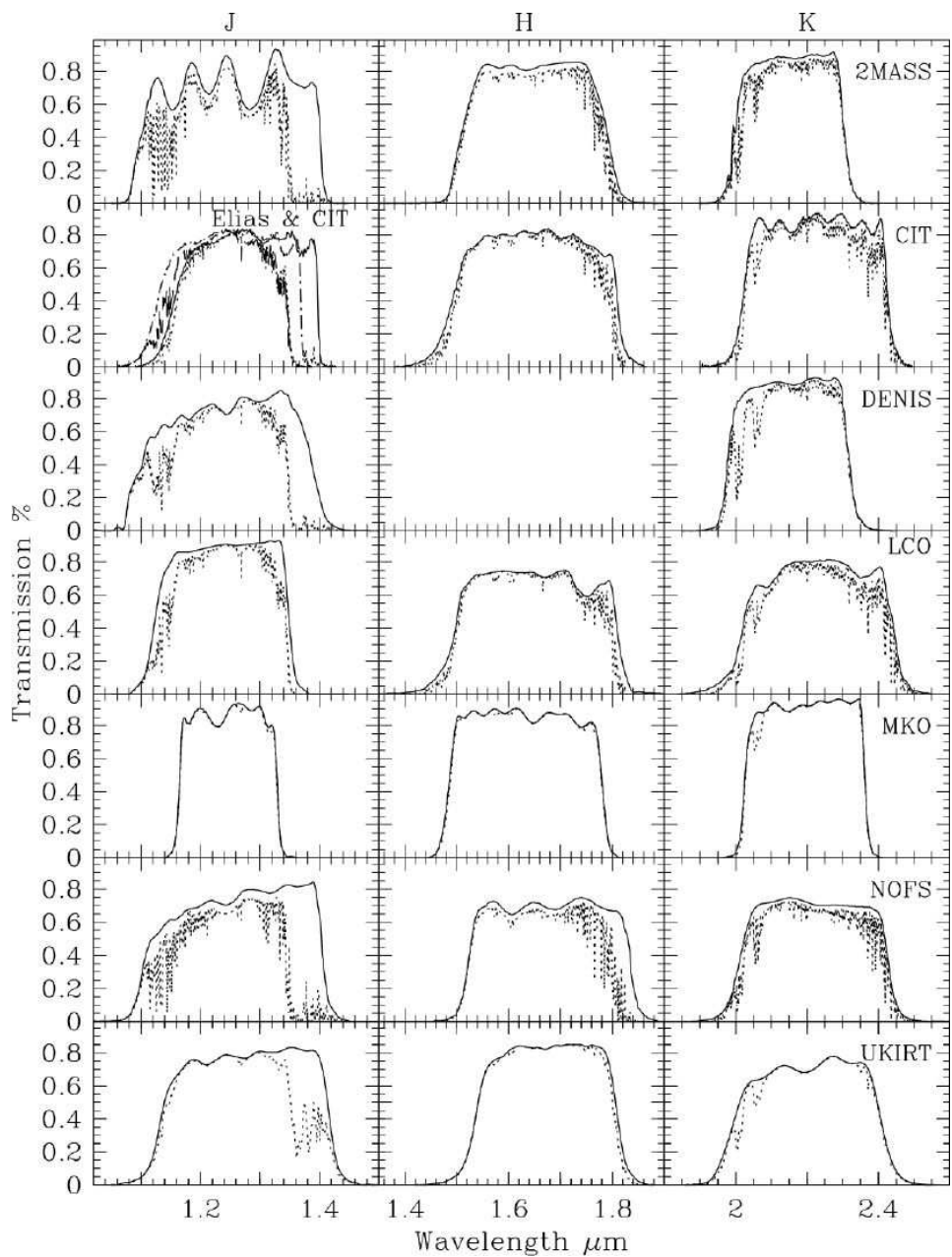}
\caption{ Filter profiles from \citet{ste04} illustrating
the variations in filter profiles of commonly used photometric
systems.  The atmospheric transmission is shown by the dotted lines.
}
\end{figure}

\clearpage
\begin{figure}
\plotone{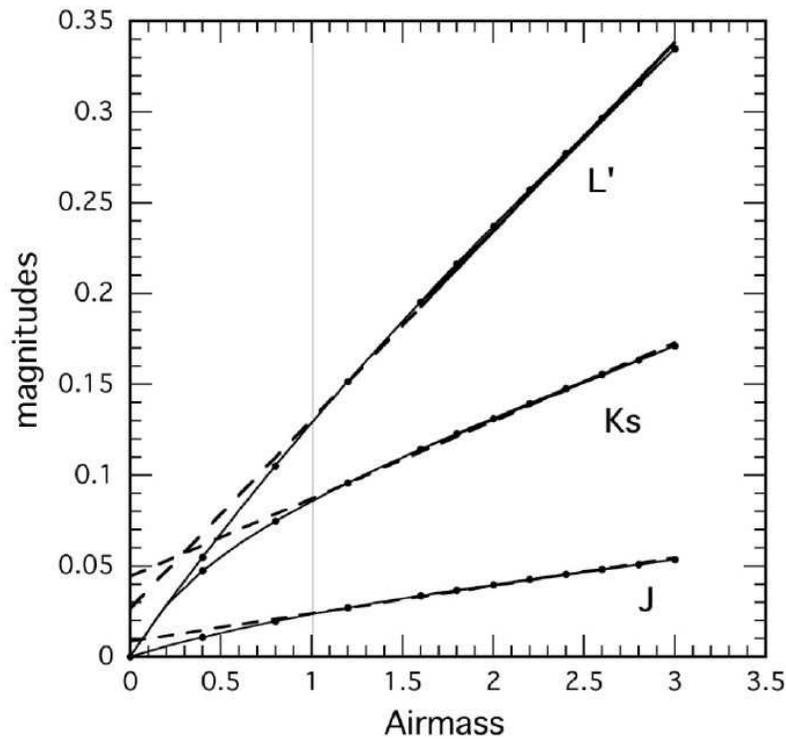}
\caption{ MKO-NIR filter extinction curves showing the linear
extrapolation to zero airmass (dashed line) for selected filters 
(adapted from \citet{tok02}.  The solid curves show the
nonlinear shape of the extinction curve between 0 and 1 airmass; 
thus extrapolation to zero airmass requires knowledge of the water
vapor content of atmosphere above the observatory and stable
conditions.  Alternatively, narrowband filters such as those proposed
by Milone \& Young (2005) should be used.
}
\end{figure}

\clearpage

\end{document}